# Reply to: "Section IV. G. Mitri's cross sections" in the paper titled "Multiple scattering and scattering cross sections" [J. Acoust. Soc. Am. 143, 995 (2018)]

F.G. Mitri*

(*Email: F.G.Mitri@ieee.org)

A B S T R A C T

The aim of this note is to rebut some unsupported claims which cast suspicions on the results of the papers titled: "Extrinsic extinction cross-section in the multiple acoustic scattering by fluid particles," [J. Appl. Phys. **121**, 144904 (2017)]; and "Intrinsic acoustical cross sections in the multiple scattering by a pair of rigid cylindrical particles in 2D," [J. Phys. D: Appl. Phys. **50**, 325601 (2017)]. It is important to emphasize that the results presented in these works are correct and valid, in complete agreement with the basic physical law of energy conservation. Moreover, scientific convincing results in peer-reviewed publications should be supported by illustrative examples using rigorous methods as well as computational analyses and solid *conclusive* evidence, instead of raising mere suspicions with baseless speculations.

The aim of this Reply is to refute some unsupported statements given in Section **IV.G.** of [J. Acoust. Soc. Am. **143**, 995 (2018)], which cast suspicions on the correctness of the results of the published peer-reviewed papers [J. Appl. Phys. **121**, 144904 (2017); J. Phys. D: Appl. Phys. **50**, 325601 (2017)]. Moreover, comments such as: "*No definitive answer has been given.*" and "*Further analysis is required.*" lead the reader astray, and suggest that the results presented in [J. Acoust. Soc. Am. **143**, 995 (2018)] are in progress, but by no means conclusive.

To alleviate any possible confusion that may occur to the reader, it is of particular importance to note the following points:

1- The "*intrinsic*" terminology used in the papers [1-3] is entirely suitable and adequate to characterize the local cross-sections presented therein, and make a clear distinction from the *extrinsic* ones defined originally in Ref.[2].

2- Concerning the derivation for the extrinsic cross-sections in [2], there is absolutely nothing "suspect" by using the asymptotic form of the cylindrical Bessel function in the mathematical derivation of the cross-sections.

   a. Eq.(20) in Ref. [2] is totally adequate to express the incident pressure field in the far-field limit and use it to derive the analytical expressions for the extrinsic absorption (Eq.(18) in [2]) and extinction (Eq.(19) in [2]) cross-sections based on energy conservation, as this physical law provides the gold standard test for the verification and adequate validation of the results. Moreover, it is basic knowledge that the solutions of equations of steady-state oscillations describing waves with sources at infinity (i.e., *plane waves*) do not satisfy radiation conditions [see: *Radiation conditions - Encyclopedia* *of Mathematics*]. This, however, does not create any singularity or difficulty in the evaluation of the cross-sections since the integrals depend on the time-averaged *product* of the pressure and velocity terms [which are convergent series], but not solely the expression of the incident pressure field. The reader is cautioned that similar expressions for the incident pressure field in the far-field limit in cylindrical coordinates have been used previously when deriving the analytical equations for the acoustical cross-sections for a single particle (see Eq.(3) in [4]; Eq.(4) in [5]) in a non-viscous fluid, and for a scatterer in an elastic matrix [6], where the corresponding results are in complete agreement with energy conservation. They have been also used in the evaluation of the radiation force [7] and torque [8] for circular and elliptical cylinders [9-11]. Another analysis generalizing the "optical theorem" for the case of arbitrary-shaped beams in spherical coordinates [12] has also used equivalent asymptotic expressions. The electromagnetic counterpart in cylindrical coordinates has been also developed [13].

   b. As an example for the case of a pair of rigid (sound impenetrable) cylinders, the reader is referred to panels (a)-(c) of Fig. 4 in Ref.[2]. The plot in panel (a) (corresponding to the dimensionless extrinsic scattering cross-section) is computed using Eq.(26) of Ref. [2], whereas the one displayed in panel (b) (corresponding to the dimensionless extrinsic extinction cross-section) is computed using Eq.(28) of Ref. [2], which is presumably "suspect" according to [J. Acoust. Soc. Am. **143**, 995 (2018)]. As shown in Fig. 4 of [2], the plots in panels (a) and (b) are *equal* such that the dimensionless extrinsic absorption cross-section is zero, which is expected since the cylinders are sound impenetrable. These results are in complete agreement with the law of energy conservation, and totally attest and confirm the accuracy and validity of Eq.(28) without any suspicion, i.e., Eq.(28) equals Eq.(26) for a pair of lossless cylinders.

   c. Another verification test was performed using *independent data* from the paper [14], and Fig. 3 in Ref.[2] using Eq.(26) of Ref.[2] displays the results which are in total agreement with those published previously in [14]. How can Eq.(28) [or Eq.(26)] in Ref. [2] be "suspect" when they agree completely with the previous results of Ref. [14]? Without a doubt, this additional test further verifies the accuracy, validity and correctness of the results presented in Ref. [2].

3- Concerning the "*intrinsic*" cross-sections [1], which reveal characteristic properties entirely connected with the probed particle under consideration [while the extrinsic ones are related to the global properties of the cluster], further comments and additional computations based on the expressions given by Eqs.(22)-(30) in Ref. [1] are presented.

   a. Firstly, since the scattering coefficients are coupled, i.e., "… [they] *are a function of both objects*..." p. 5 in [1], it is obviously implied that multiple interference effects are incorporated in them, as shown by Eqs.(31) and (32) of Ref. [1] and acknowledged therein, with the dependence of the series on the cylindrical Hankel function of the first kind with argument $kd$. The intrinsic scattering cross-section provides quantitative information on the scattering properties of the probed object exclusively (but in the presence of the second [or other] particle) without *additional* multiple interference reverberating effects, described by the third factor in Eq.(28) of Ref. [1], with the dependence of the series on the cylindrical Bessel function of the first kind with argument *kd*.

b. Secondly, consider the case of two rigid sound impenetrable cylinders of same radii. The panels in Fig. 1 (below) show the plots for the extrinsic (superscript "e") and intrinsic (superscript "i") energy efficiencies. Comparison of panel (a) with (b), panel (d) with (e) and panel (g) with (h) clearly shows that the corresponding plots are identical. Moreover, panels (f) and (i) in Fig. 1 clearly show that the *intrinsic* efficiencies for the rigid cylinders are zero. This is in complete agreement with the energy conservation law, stating that the extinction cross-section equals the scattering cross-section in the absence of absorption. In other words, when both objects are lossless, the extrinsic and intrinsic absorption cross-sections (or efficiencies) are zero, while the extinction and scattering efficiencies are equal as expected. Therefore, in contrast with the claim in [J. Acoust. Soc. Am. **143**, 995 (2018)], the results presented in [1-3] and those shown in Fig. 1, demonstrate total physical validity based on energy conservation, and lack of any suspicion. Other examples for non-viscous fluids and elastic cylinders were considered and showed similar results (i.e. zero absorption efficiency), but are not presented here for brevity.

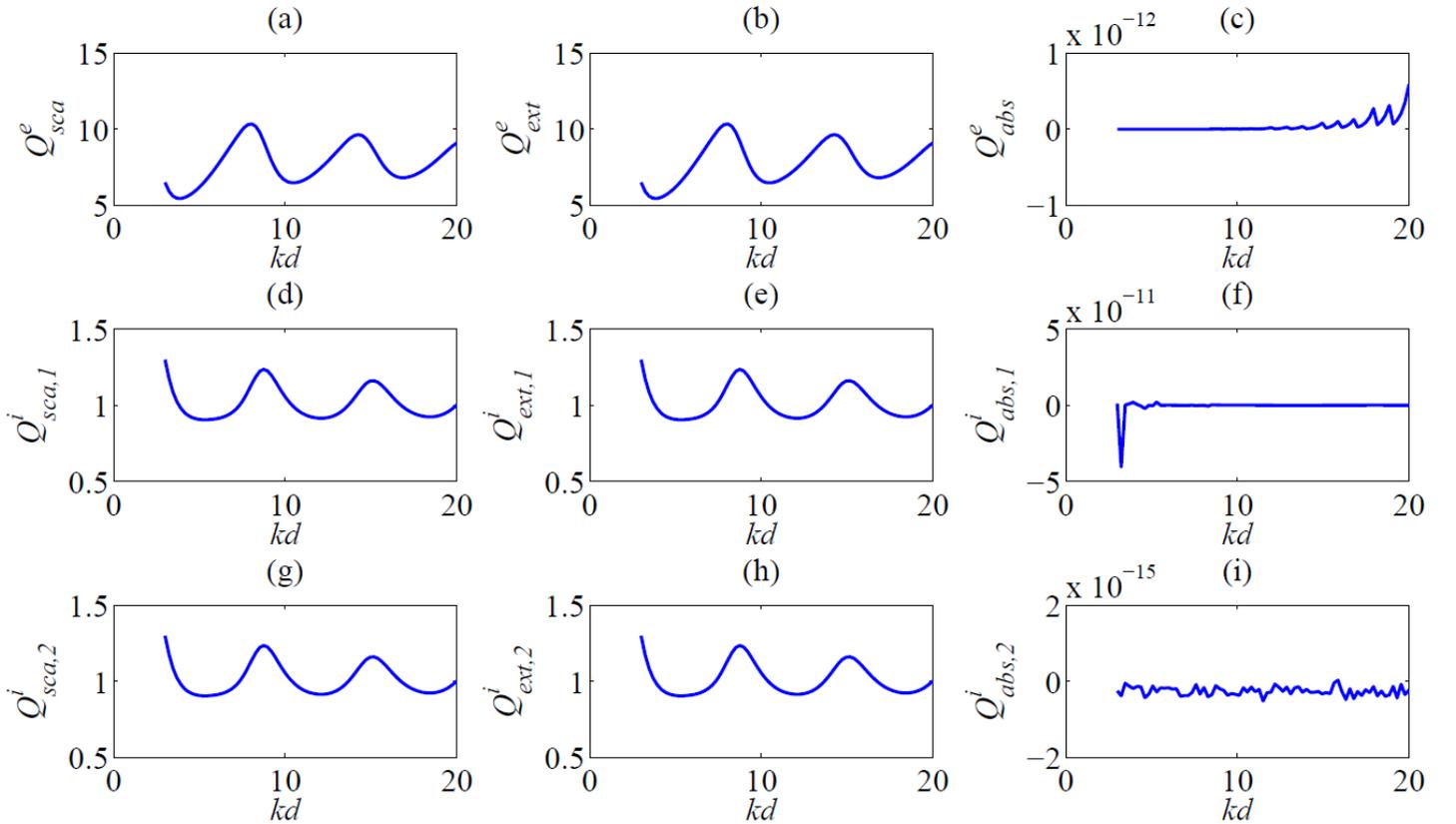

**Fig. 1.** Panels (a)-(c) correspond to the extrinsic cross-sections, while panels (d)-(i) correspond to the intrinsic ones for the rigid cylinder pair having $ka = 1$ and $kb = 1$ at $\alpha = 45°$.

c. Thirdly, consider the case of two fluid (i.e., liquid) cylinders made of the same material ($\rho_1 = \rho_2 = 656$ kg/m$^3$; speed of sound for the compressional waves, $c_1 = c_2 = 1405$ m/s; dimensionless attenuation coefficient satisfying energy conservation [15] $\gamma_1 = 10^{-3}$, $\gamma_2 = 0$); the first cylinder (labeled 1 in [1, 2]) is viscous, while the second is not ($\gamma_2 = 0$). The panels in Fig. 2 (below) show the corresponding plots for the extrinsic (superscript "e" – first row) and intrinsic energy efficiencies (superscript "i" – rows 2 and 3 for objects 1 and 2, respectively).

The results in Fig. 2 show that when a multiple scattering system of two particles is considered in which one object is sound-absorptive (cylinder 1) while the second (cylinder 2) is not, non-vanishing *extrinsic*/global absorption efficiency for both objects [based on Eq.(29) in [1]; and panel (c) in Fig. 2] in addition to a non-zero *intrinsic*/local absorption efficiency for the first object [based on Eq.(23) in [1]; and panel (f) in Fig. 2] can be defined consecutively and evaluated numerically. However, **the intrinsic/local absorption efficiency for the second object vanishes** [based on Eq.(26) in [1]; and demonstrated in panel (i) of Fig. 2]. Thus, the example results in Fig. 2 clearly show that the *intrinsic* cross-sections/efficiencies would lead to quantitative assessment of the local properties of each of the scatterers in an acoustically-interacting multiple scattering system [1], in agreement with the law of energy conservation. For the viscous cylinder 1, a quantifiable (non-zero) intrinsic absorption efficiency can be computed, while for the non-viscous cylinder 2, the intrinsic absorption efficiency vanishes as demonstrated in panel (i). Clearly, such meaningful information related to cylinder 2 cannot be obtained from the plot of the extrinsic/global absorption efficiency shown in panel (c) of Fig. 2. This suggests that the intrinsic cross-sections provide local properties for each particle in the multiple interacting system composed of many particles. In addition, the results in Fig. 2 are physical and in total agreement with energy conservation as confirmed by the computational results obtained using the rigorous partial-wave series expansion method in cylindrical coordinates.

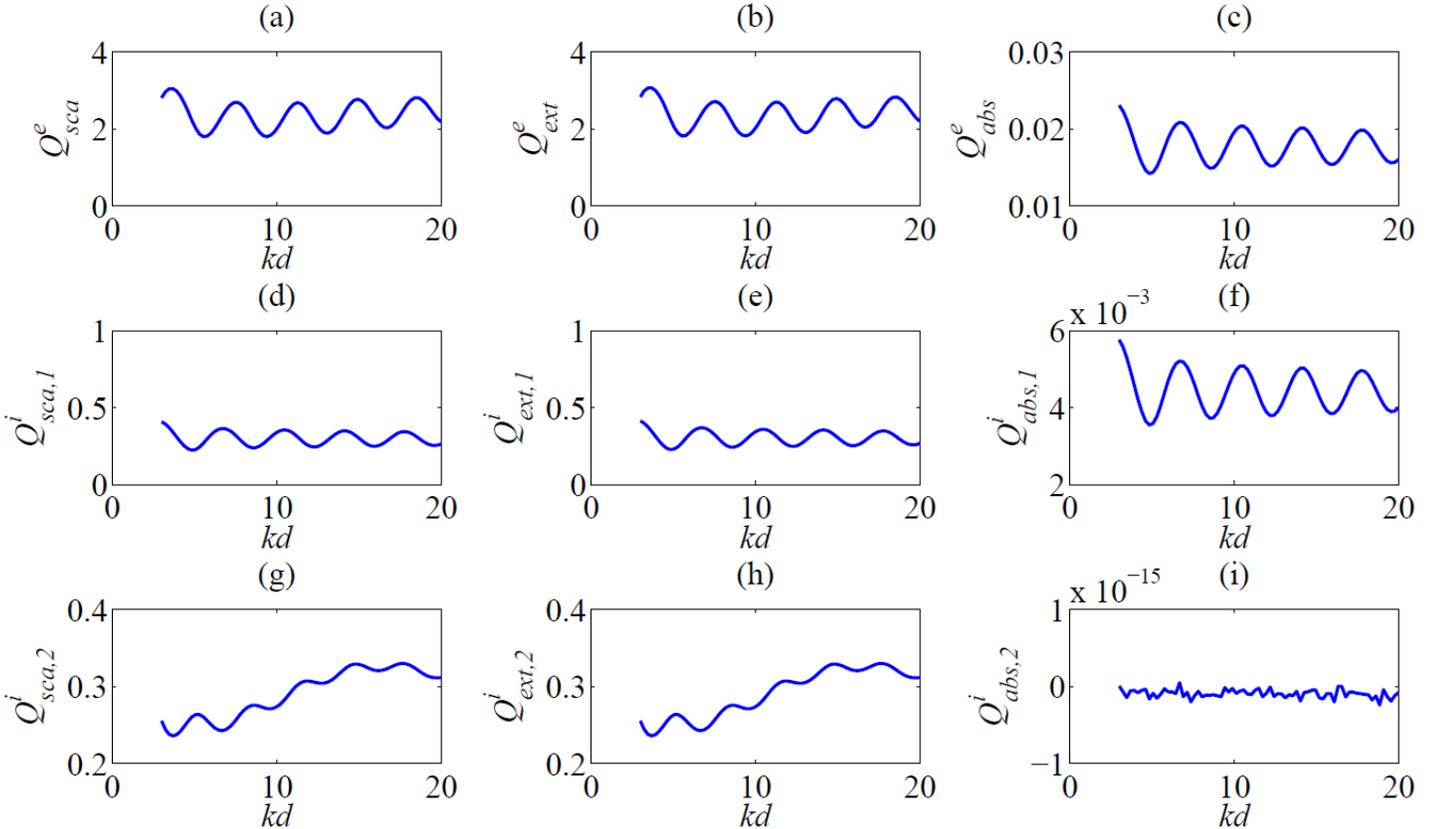

**Fig. 2.** Panels (a)-(c) correspond to the extrinsic cross-sections, while panels (d)-(i) correspond to the intrinsic ones for the viscous fluid cylinder pair having $ka = 1$ and $kb = 1$ at $\alpha = 45°$.

4- It is important to note that the method based on the adequate determination of an *effective incident field* used in [1-3] allows to obtain meaningful information related to the following physical observables, such as

   **i)** the extrinsic and intrinsic cross-sections [1-3],
   **ii)** the acoustic radiation forces [16, 17] (and their electromagnetic counterpart [18]),
   **iii)** and the acoustic radiation torques [19].

   The "*effective field*" terminology (see Eq.(3.2) in [20] and [21]) has been initially introduced in the multiple scattering of waves as an "*external field*" (Eq.(5) in [22]) acting on the probed particle. In the recent literature, it is also known as the "*excitation/exciting field*" (Ch. 7 in [23]), which is the superposition of the incident primary field and the field scattered by all the other neighboring particles in the system (excluding the field scattered by the probed particle itself). The acoustic radiation stress on the probed particle [which depends on both the *effective* incident field and the scattered field] is a zero-divergence tensor in a non-viscous host fluid. By doing so, the problem in hand is reduced to the *single scatterer configuration* with an effective field incident upon its surface. Subsequently, the method for determining the related cross-sections, radiation forces and torques is based on a *far-field scattering approach* where the integration over a surface at a large radius enclosing the probed object cylinder is performed. Therefore, the far-field method gives exact results without any approximations. This is an early method recognized in electromagnetism [24] dealing with the radiation force on a spherical particle utilizing two scalar potential functions, and later extended to acoustics from the standpoint of radiation force [25] and torque [26] theories.

In conclusion, the works [1-3] presented correct physical results and mathematical expressions for the cross-sections based on rigorous analyses in complete agreement with the basic law of energy conservation and without any suspicion. The derived mathematical expressions presented in [1-3] are entirely suitable for predicting the intrinsic cross-sections for the individual interacting particles, as well as the extrinsic ones for the cluster, respectively. The points discussed in this Reply placing the works [1-3] in the appropriate context and aiming to alleviate potential confusion to the reader, should be helpful to the scientific community interested in the topic of acoustic extinction and other subjects dealing with acoustic radiation forces [16] and torques [19] in multiple interacting systems.

**REFERENCES**

[1] F. G. Mitri, "Intrinsic acoustical cross sections in the multiple scattering by a pair of rigid cylindrical particles in 2D," *J. Phys. D: Appl. Phys.,* vol. 50, no. 32, art. no. 325601, 2017.
[2] F. G. Mitri, "Extrinsic extinction cross-section in the multiple acoustic scattering by fluid particles," *J. Appl. Phys.,* vol. 121, no. 14, art. no. 144904, 2017.
[3] F. G. Mitri, "Extrinsic and intrinsic acoustical cross-sections for a viscous fluid particle near a planar rigid boundary: circular cylinder example," *J. Phys. Communications,* vol. 1, no. 5, art. no. 055015, 2017.
[4] F. G. Mitri, "Optical theorem for two-dimensional (2D) scalar monochromatic acoustical beams in cylindrical coordinates," *Ultrasonics,* vol. 62, pp. 20-26, 2015.
[5] F. G. Mitri, "Extended optical theorem for scalar monochromatic acoustical beams of arbitrary wavefront in cylindrical coordinates," *Ultrasonics,* vol. 67, pp. 129 - 135, 2016.
[6] F. G. Mitri, "Extinction efficiency of "elastic–sheet" beams by a cylindrical (viscous) fluid inclusion embedded in an elastic medium and mode conversion—Examples of nonparaxial Gaussian and Airy beams," *J. Appl. Phys.,* vol. 120, no. 14, art. no. 144902, 2016.
[7] F. G. Mitri, "Interaction of an acoustical 2D-beam with an elastic cylinder with arbitrary location in a non-viscous fluid," *Ultrasonics,* vol. 62, pp. 244–252, 2015.
[8] F. G. Mitri, "Acoustic radiation force and spin torque on a viscoelastic cylinder in a quasi-Gaussian cylindrically-focused beam with arbitrary incidence in a non-viscous fluid," *Wave Motion,* vol. 66, pp. 31-44, 2016.
[9] F. G. Mitri, "Acoustic radiation force on a rigid elliptical cylinder in plane (quasi)standing waves," *J. Appl. Phys.,* vol. 118, no. 21, art. no. 214903, 2015.
[10] F. G. Mitri, "Acoustic backscattering and radiation force on a rigid elliptical cylinder in plane progressive waves," *Ultrasonics,* vol. 66, pp. 27-33, 2016.
[11] F. G. Mitri, "Radiation forces and torque on a rigid elliptical cylinder in acoustical plane progressive and (quasi)standing waves with arbitrary incidence," *Phys. Fluids,* vol. 28, no. 7, p. 077104, 2016.
[12] F. G. Mitri and G. T. Silva, "Generalization of the extended optical theorem for scalar arbitrary-shape acoustical beams in spherical coordinates," *Phys. Rev. E,* vol. 90, no. 5, p. 053204, 2014.
[13] F. G. Mitri, "Generalization of the optical theorem for monochromatic electromagnetic beams of arbitrary wavefront in cylindrical coordinates," *J. Quant. Spectr. Rad. Transfer,* vol. 166, pp. 81-92, 2015.
[14] J. A. Roumeliotis, A.-G. P. Ziotopoulos, and G. C. Kokkorakis, "Acoustic scattering by a circular cylinder parallel with another of small radius," *J. Acoust. Soc. Am.,* vol. 109, no. 3, pp. 870-877, 2001.
[15] F. G. Mitri and Z. E. A. Fellah, "Physical constraints on the non-dimensional absorption coefficients of compressional and shear waves for viscoelastic cylinders," *Ultrasonics,* vol. 74, pp. 233-240, 2017.
[16] F. G. Mitri, "Acoustic attraction, repulsion and radiation force cancellation on a pair of rigid particles with arbitrary cross-sections in 2D: Circular cylinders example," *Ann. Phys.* vol. 386, pp. 1-14, 2017.
[17] F. G. Mitri, "Acoustic radiation force on a cylindrical particle near a planar rigid boundary," *J. Phys. Communications,* https://doi.org/10.1088/2399-6528/aab1109, 2018.
[18] F. G. Mitri, "Pushing, pulling and electromagnetic radiation force cloaking by a pair of conducting cylindrical particles," *J. Quant. Spectr. Rad. Transfer,* vol. 206, pp. 142-150, 2018.
[19] F. G. Mitri, "Acoustic radiation torques on a pair of fluid viscous cylindrical particles with arbitrary cross-sections: Circular cylinders example," *J. Appl. Phys.,* vol. 121, no. 14, art. no. 144901, 2017.
[20] M. Lax, "Multiple scattering of waves," *Rev. Mod. Phys.,* vol. 23, no. 4, pp. 287-310, 1951.
[21] M. Lax, "Multiple scattering of waves. II. The effective field in dense systems," *Phys. Rev.,* vol. 85, no. 4, pp. 621-629, 1952.
[22] L. L. Foldy, "The multiple scattering of waves. I. General theory of isotropic scattering by randomly distributed scatterers," *Phys. Rev.,* vol. 67, no. 3-4, pp. 107-119, 1945.
[23] L. Tsang, J. A. Kong, K. H. Ding, and C. O. Ao, *Scattering of Electromagnetic Waves, Numerical Simulations*: Wiley, 2001.
[24] P. Debye, "Der lichtdruck auf kugeln von beliebigem material," *Ann. der Phys.,* vol. 335, no. 11, pp. 57-136, 1909.
[25] P. J. Westervelt, "Acoustic radiation pressure," *J. Acoust. Soc. Am.,* vol. 29, no. 1, pp. 26-29, 1957.
[26] G. Maidanik, "Torques due to acoustical radiation pressure," *J. Acoust. Soc. Am.,* vol. 30, no. 7, pp. 620-623, 1958.

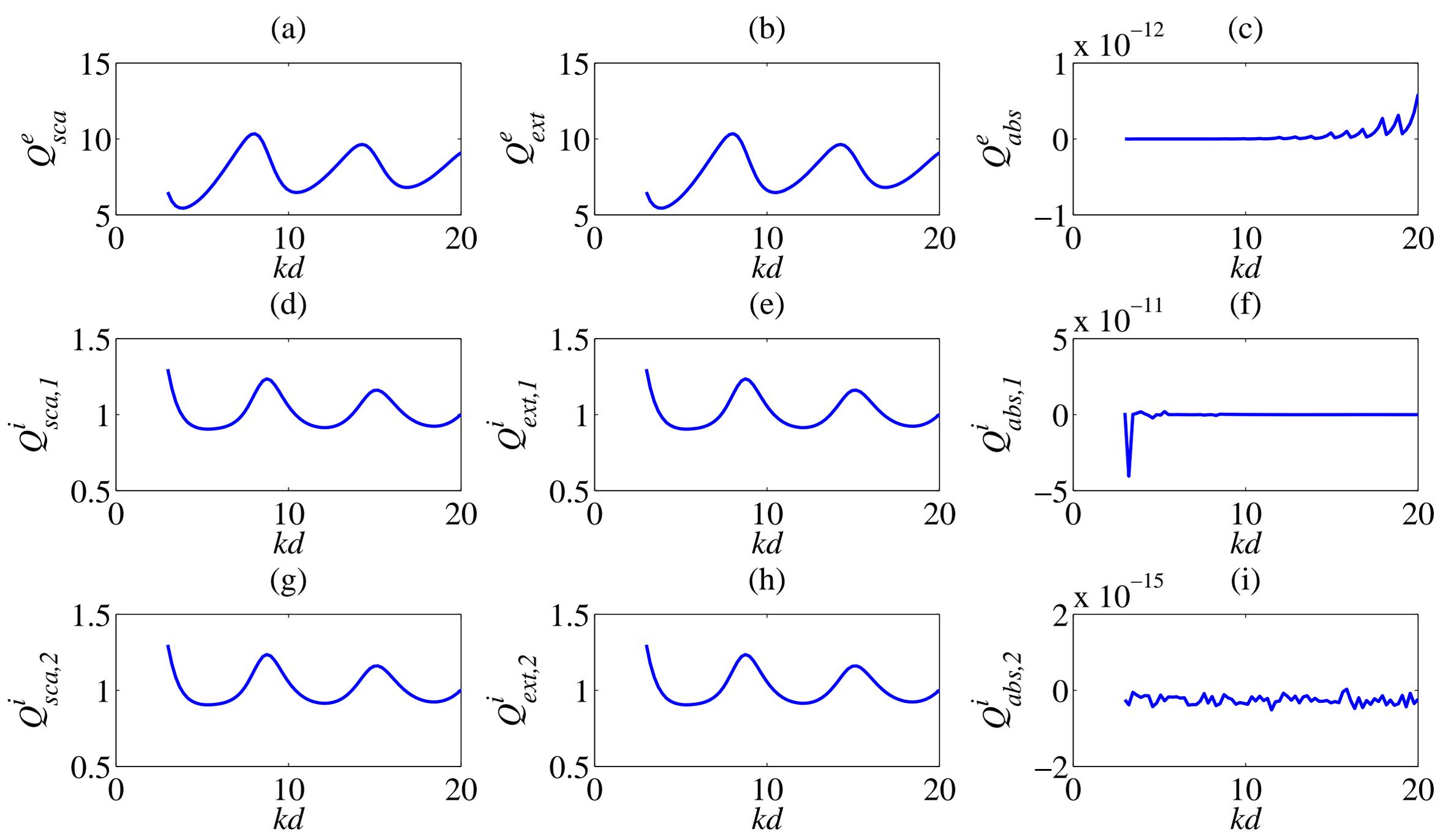

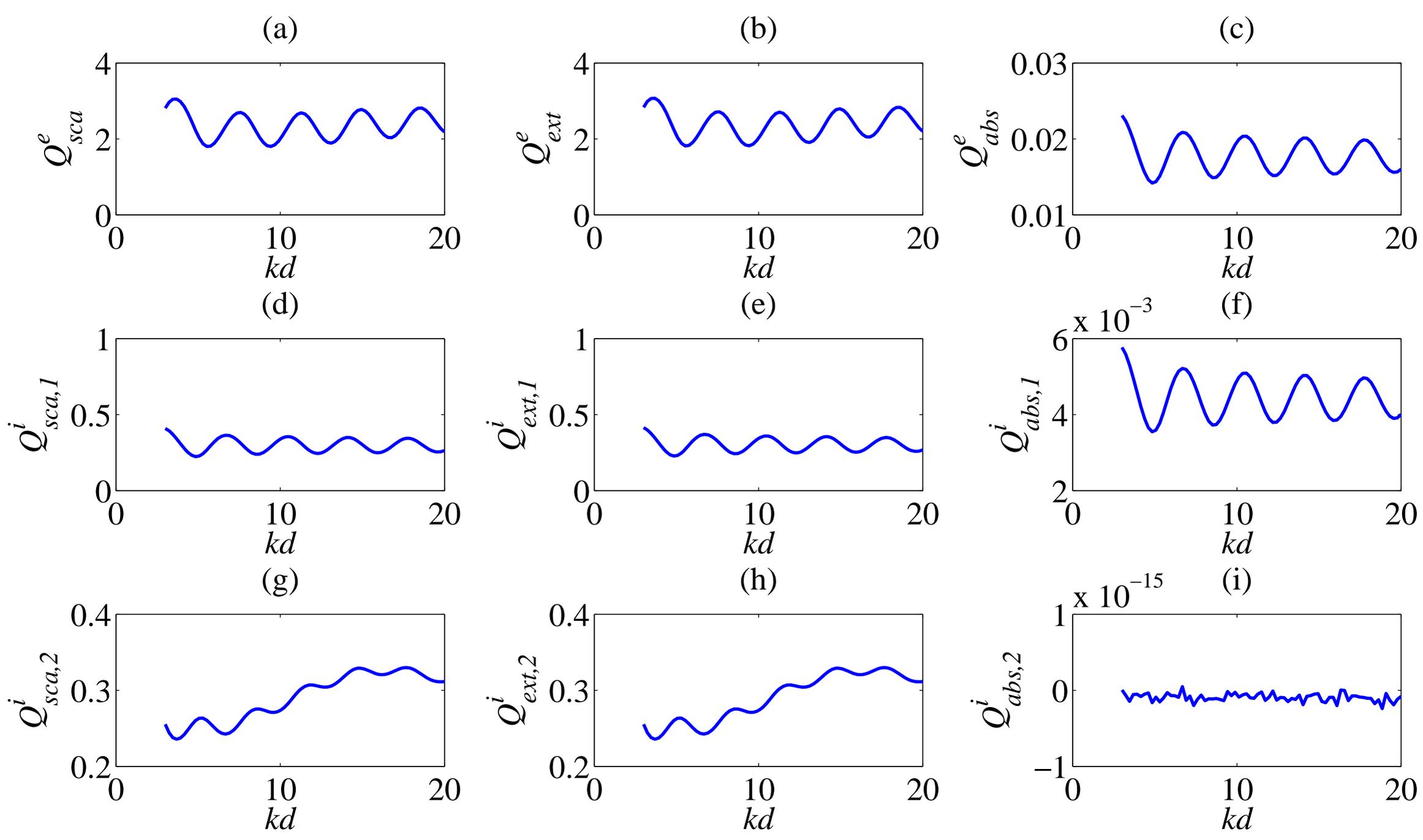